\documentstyle[epsfig,twocolumn,aps,prl]{revtex}

\begin{document}
\draft
\title{Two-pion correlations in Au+Au collisions at 10.8~GeV/c per nucleon}

\author{
J.~Barrette$^{(4)}$, R.~Bellwied$^{(8)}$, S.~Bennett$^{(8)}$, 
R.~Bersch$^{(5)}$, P.~Braun-Munzinger$^{(5),}$\cite{gsi}, 
W.~C.~Chang$^{(5)}$, W.~E.~Cleland$^{(6)}$, 
J.D.~Cole$^{(3)}$, T.~M.~Cormier$^{(8)}$, G.~David$^{(1)}$, 
J.~Dee$^{(5)}$, O.~Dietzsch$^{(7)}$, M.W.~Drigert$^{(3)}$, S.~Gilbert$^{(4)}$, 
J.~R.~Hall$^{(8)}$, T.~K.~Hemmick$^{(5)}$, N.~Herrmann$^{(2)}$, 
B.~Hong$^{(5),}$\cite{gsi}, 
C.~L.~Jiang$^{(5)}$, S.C.~Johnson$^{(5)}$, Y.~Kwon$^{(5)}$, 
R.~Lacasse$^{(4)}$, 
A.~Lukaszew$^{(8)}$, Q.~Li$^{(8)}$, T.~W.~Ludlam$^{(1)}$, S.~McCorkle$^{(1)}$,
S.~K.~Mark$^{(4)}$, R.~Matheus$^{(8)}$, D.~Mi\'skowiec$^{(5),}$\cite{gsi}, 
E.~O'Brien$^{(1)}$, 
S.~Panitkin$^{(5)}$, T.~Piazza$^{(5)}$, M.~Pollack$^{(5)}$, 
C.~Pruneau$^{(8)}$, M.~N.~Rao$^{(5)}$, 
M.~Rosati$^{(4)}$, 
N.~C.~daSilva$^{(7)}$, 
S.~Sedykh$^{(5)}$, 
U.~Sonnadara$^{(6)}$, 
J.~Stachel$^{(5),}$\cite{heidelberg}, 
E.~M.~Takagui$^{(7)}$, M.~Trzaska$^{(5)}$, 
S.~Voloshin$^{(6),}$\cite{heidelberg}, 
T.~Vongpaseuth$^{(5)}$, G.~Wang$^{(4)}$, 
J.~P.~Wessels$^{(5),}$\cite{heidelberg}, 
C.~L.~Woody$^{(1)}$, N.~Xu$^{(5)}$, 
Y.~Zhang$^{(5)}$, C.~Zou$^{(5)}$
\begin{center}
(E877 Collaboration)
\end{center}
}

\address{
$^{(1)}$ Brookhaven National Laboratory, Upton, NY 11973\\ 
$^{(2)}$ Gesellschaft f\"ur Schwerionenforschung, Darmstadt, Germany\\ 
$^{(3)}$ Idaho National Engineering Laboratory, Idaho Falls, ID 83415\\ 
$^{(4)}$ McGill University, Montreal, Canada\\ 
$^{(5)}$ State University of New York at Stony Brook, NY 11794\\
$^{(6)}$ University of Pittsburgh, Pittsburgh, PA 15260\\
$^{(7)}$ University of S\~ao Paulo, Brazil\\ 
$^{(8)}$ Wayne State University, Detroit, MI 48202\\
}

\maketitle

\begin{abstract}
Two-particle correlation functions for positive and negative pions 
have been measured in Au+Au collisions at 10.8~GeV/c per nucleon. 
The data were analyzed using one- and three-dimensional correlation 
functions. From the results of the three-dimensional fit the phase 
space density of pions was calculated. It is consistent with local 
thermal equilibrium. 
\end{abstract}


Bose-Einstein correlations of identical pions can be used to obtain 
information about the space-time-momentum distribution $S({\bf r},t,{\bf p})$ 
of pions at freeze-out (pion source) in a nuclear collision  
(for a review see \cite{boa90}). 
The two-pion correlation function $C(\bf p_1,p_2)$, 
defined as the ratio between 
the two-particle density and the product of single particle densities, 
shows a peak at ${\bf q} = {\bf p_2} - {\bf p_1} = 0$.
The width of this peak is inversely proportional to the size of 
the pion source. 
Analysis of $C$ as a function of different components of the relative momentum 
yields information about the source dimensions in different directions. 
The relation between $S$ and $C$ is, under realistic conditions, 
complicated by effects like 
three- (and more) particle correlations \cite{zaj87,pra93,mer95}, 
long-lived resonance decays \cite{gyu89,led92,pei92}, and the 
distortion of the single-particle spectra by the two-particle correlations 
\cite{zaj84,pei93}. 
The correlation must also be corrected for the Coulomb interaction between 
the two pions \cite{gyu79,pra86,bay96}. 

In this Letter we present results of the correlation analysis of pions 
produced in central Au+Au collisions at 10.8~GeV/c per nucleon. 
The data were taken in Fall 1993 at the AGS. 
The central trigger at the level of 10\% of the geometrical cross section 
was used. 
The apparatus allowed a simultaneous measurement 
of positive and negative pions with a momentum resolution of 
$\Delta p/p\approx$~3\%. 
With two field polarities used, the overall acceptances 
for positive and negative pion pairs were similar: 
$2<y<4$ and $0<p_t<0.5$~GeV/c with 
$\langle y \rangle=3.1$ and $\langle p_t \rangle=0.1$~GeV/c. 
The analyzed data sample consists of one million central events and 
the total number of analyzed $\pi^{+}\pi^{+}$, $\pi^{-}\pi^{-}$, and  
$\pi^{+}\pi^{-}$ pairs is 130~k, 210~k, and 340~k, 
respectively. 
The particle identification (PID) quality was tested by varying the PID cuts
in order to deliberately accept background particles. 
No significant influence on the results was observed 
implying that particle misidentification does not contribute significantly 
to the overall systematic uncertainty. 
A more detailed description of the experiment and of the data analysis 
can be found in \cite{mis96}. 

Experimentally, the correlation function $C({\bf q})$ is defined as 
the number of pion pairs in a $\bf q$-bin (signal) 
divided by the number of such pairs obtained by event mixing 
(background). 
The variables $q_{\rm out}, q_{\rm side}$, and $q_{\rm long}$, 
used in the three-dimensional `out-side-long' analysis, 
are the components of $\bf q$ in the beam rapidity frame 
$y_{\rm anal}=y_{\rm beam}=3.14$.
Here $q_{\rm long}$ is the component parallel to the beam, 
$q_{\rm side}$ is perpendicular to the beam and to the average pair momentum, 
and $q_{\rm out}$ is perpendicular to $q_{\rm long}$ and 
$q_{\rm side}$ \cite{ber88,pra90}.
In our one-dimensional analysis we analyzed two-pion correlations as 
a function of $q=|{\bf q}|$ calculated in the pair c.m. frame. 
For equal mass particles $q$ is equal to 
$Q_{\rm inv}$ which is defined as 
$\sqrt{({\bf p_2}-{\bf p_1})^2 - (E_2-E_1)^2}$.

The background pairs were weighted with the Coulomb correction factor 
to remove the effect of the Coulomb interaction between the two pions
from the correlation. 
It is calculated by taking the square of the non-relativistic wave 
function describing a particle in a Coulomb field \cite{mer,mes}:
\begin{equation}
H({\bf k},{\bf r})=
\frac{2\pi\eta}{e^{2\pi\eta}-1} \ \
\left|F\left(-i\eta;1;i k (r-\frac{{\bf r}\cdot{\bf k}}{k})\right)\right|^2 \ ,
\end{equation}
where 
$F$ is the confluent hypergeometric function, 
$\bf k$ is the asymptotic momentum of a pion in the pair c.m.s. 
($k=Q_{\rm inv}/2$), and 
$\bf r$ is the relative distance of the two pions at freeze-out. 
The relative velocity enters through 
$\eta=Z_1 Z_2 m_\pi c \alpha / Q_{\rm inv}$ 
where $m_\pi$ is the pion mass and $\alpha$ is the fine structure constant.  
For pions emitted from a point-like source 
$F$ is equal to unity and 
the Coulomb correction becomes 
the well known Gamow factor \cite{gyu79,mer} (dot-dashed line in Fig.~1). 
For a finite size source, the Coulomb correction factor 
can be obtained numerically by averaging $H({\bf k},{\bf r})$ 
over ${\bf r}$. 
The correction $H$ calculated this way for a Gaussian source 
with $\sigma_x=\sigma_y=\sigma_z=5$~fm and $\sigma_t=0$ 
is shown as a dashed curve in Fig.~1. 
In addition, 
since it would be incorrect to divide data 
which have been measured with a finite resolution $\Delta p$ 
by a correction calculated assuming $\Delta p$=0, we folded $H$ 
with the momentum resolution of the spectrometer. 
The resulting correction $H_{\Delta p}$ is shown as a solid line in Fig.~1. 
\begin{figure}
\psfig{figure=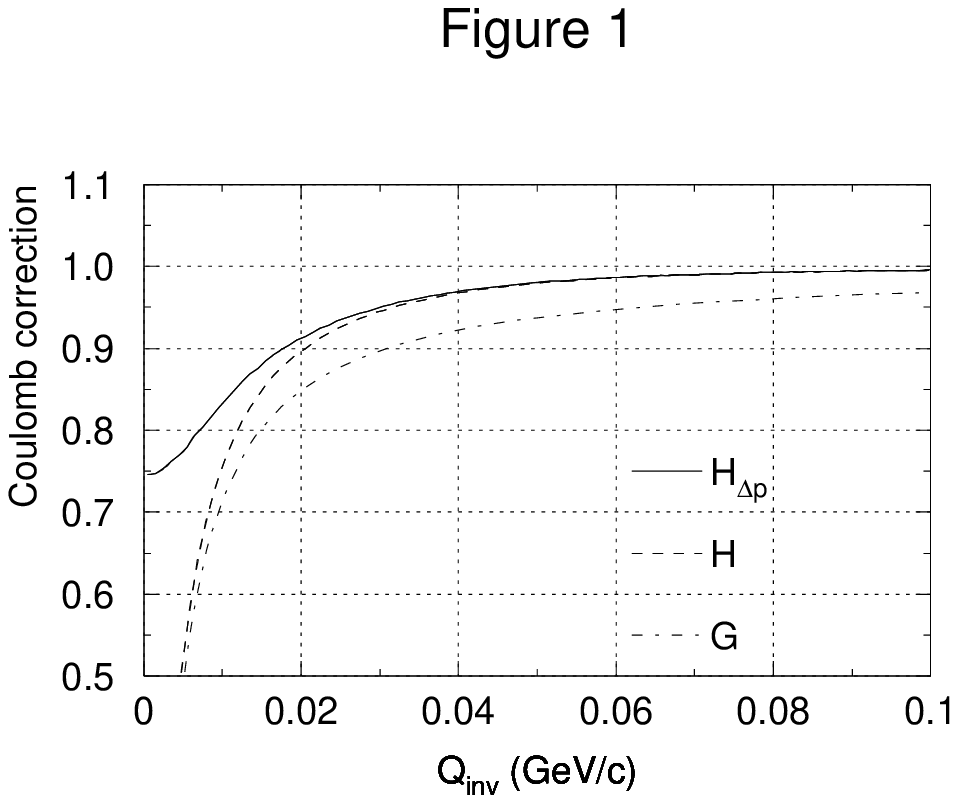,height=7.5cm,width=9cm}
\caption{
Coulomb correction factor for two-pion correlation functions. 
The dot-dashed line $G$ represents a correction assuming a point-like 
pion source (Gamow factor). 
The dashed line $H$ was calculated for a Gaussian source 
with a realistic size. 
The same curve folded with the spectrometer momentum resolution is shown 
as a solid line. 
This is the correction used in the data analysis.}
\end{figure}

Since the $\pi^+ \pi^-$ correlation is dominated by the mutual Coulomb 
interaction, we can use it to test the quality of the Coulomb correction. 
This correlation function, uncorrected, corrected by the Gamow, 
and by $H_{\Delta p}$ is shown in the right hand panels of Fig.~2. 
The properly corrected correlation should be equal to unity. 
The Gamow correction is obviously inappropriate. 
This is in agreement with several theoretical studies 
\cite{pra86,bow91,biy95} 
as well as previous experimental observations \cite{vos94,alb95}. 
The by $H_{\Delta p}$ corrected $\pi^+\pi^-$ correlation is close 
to unity. 
We therefore use the same method to correct the 
$\pi^+ \pi^+$ and $\pi^- \pi^-$ correlation functions. 
\begin{figure}[t!]
\psfig{figure=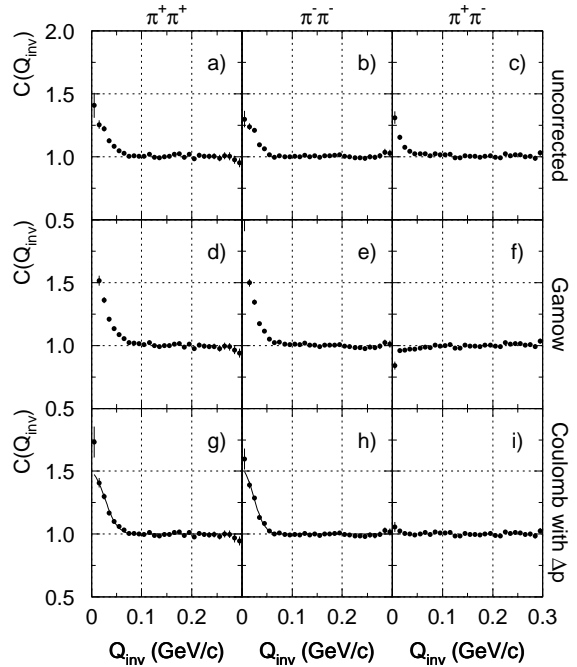,height=11.3cm,width=9cm}
\caption{
One-dimensional $\pi^+\pi^+, \pi^-\pi^-$, and $\pi^+\pi^-$ 
correlation functions. 
Presented are from top to bottom: raw and Gamow corrected correlations, 
and correlations corrected by $H_{\Delta p}$. 
The solid line is the Gaussian fit.}
\end{figure}
The Gamow correction would lead to a significantly different shape 
of the corrected correlation (see Fig.~2 and Table~I). 

In order to test possible distortions of the correlation function 
introduced by the apparatus and the data analysis 
we performed a Monte Carlo simulation 
in which 
we generated events with realistic particle multiplicities and momenta, 
and propagated the particles through our setup using GEANT \cite{geant}. 
The simulated detector response was 
used as an input to the data analysis code. 
The obtained correlation functions were equal to unity as expected 
for a simulation without two-particle interactions. 
Subsequently, we generated events in which positive pions were correlated. 
The obtained correlation functions were fitted 
and the fit results were compared to the parameters of the original 
correlation. 
The observed 3-10\% and 20\% reductions of the radius parameters
and the correlation strength (see below for definitions), 
respectively,  
are consistent with the broadening of the correlation peak expected 
from the finite momentum resolution of the spectrometer. 

The experimental, Coulomb corrected one-dimensional 
$\pi^+\pi^+$ and $\pi^-\pi^-$ correlations were parametrized by 
\begin{equation}
\label{one-dim-fit}
C(Q_{\rm inv})=1+\lambda \exp {}(-R^2 Q_{\rm inv}{}^2)
\end{equation}
with $\hbar=1$. 
The results of a maximum likelihood fit are shown as the solid line 
in Fig.~2. 
The fit parameters are given in Table~I. 
Fitting Coulomb uncorrected correlations gives the same $R$ but a 
lower $\lambda$ value of 0.32. 
We estimate the uncertainty of $\lambda$ because of the assumptions entering 
the Coulomb correction to be half of the difference, i.e. 18\%. 

In the `out-side-long' analysis, 
which was performed in the beam rapidity frame $y_{\rm anal}=3.14$, 
the three-dimensional correlation functions were fitted by 
\begin{eqnarray}
\label{three-dim-fit}
\nonumber
C(q_{\rm out},q_{\rm side},q_{\rm long})&=&
1+\lambda \exp(-R_{\rm out}{}^2q_{\rm out}{}^2-
R_{\rm side}{}^2 q_{\rm side}{}^2 +\\
&&-R_{\rm long}{}^2 q_{\rm long}{}^2 - 
2|R_{\rm ol}| R_{\rm ol} q_{\rm out}q_{\rm long}) \ .
\end{eqnarray}
The projections of the correlation and of the fit, 
obtained by averaging over the two other components of $\bf q$ 
in the range from -50~MeV/c to 50~MeV/c 
with a weighting factor equal to the number of background pairs, 
are shown in Fig.~3. 
\begin{figure}[b!]
\psfig{figure=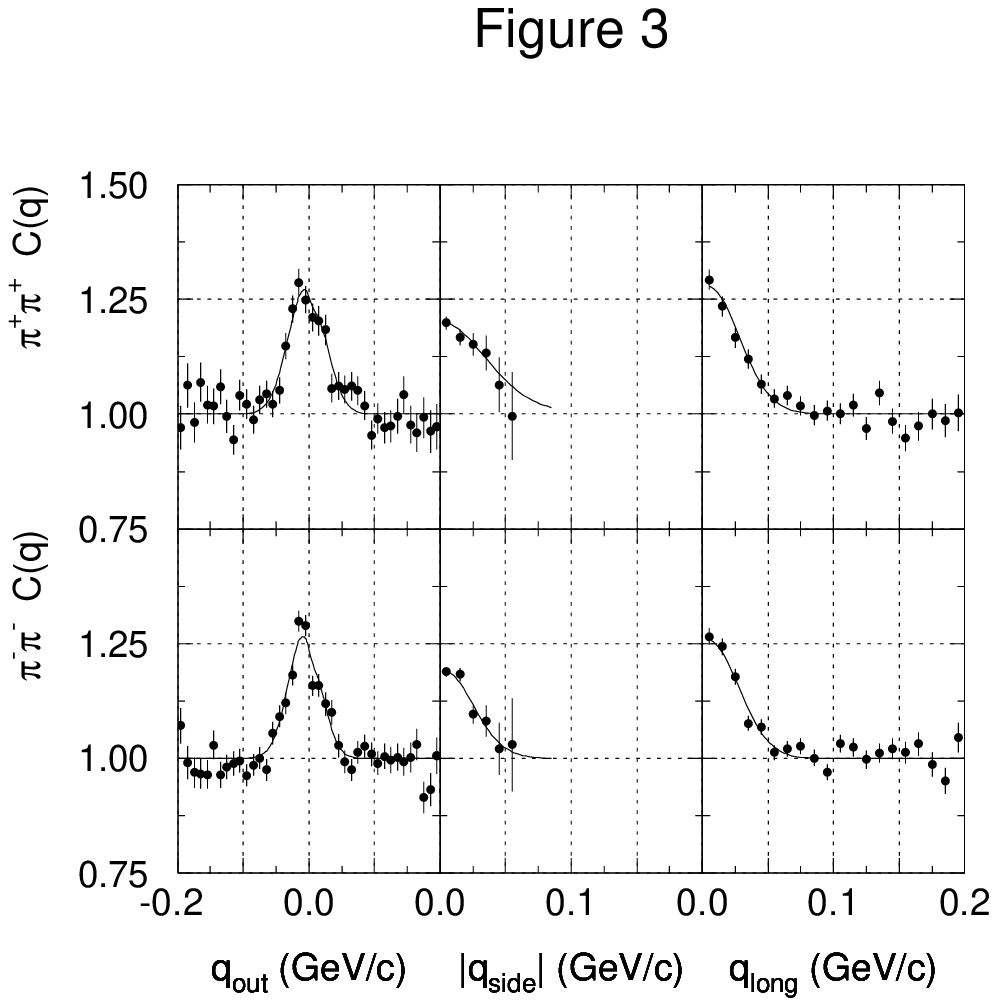,height=9cm,width=9cm}
\caption{
Projections of the three-dimensional $\pi^+\pi^+$ and $\pi^-\pi^-$ 
correlation functions. 
Solid lines represent projections of the fit.}
\end{figure}
The particles were labeled such that $q_{\rm long}$$>$0. 
The asymmetric shape of the projection $C(q_{\rm out})$, 
well reproduced by the fit, 
is caused by the acceptance and the finite width of the projection window. 
The parameters obtained from the fit are given in Table~II. 
The values were 
corrected for the momentum resolution of the spectrometer using the 
results of the Monte Carlo simulation. 

Similarly as in the one-dimensional fit, the $\pi^+$ and $\pi^-$ 
sources apparently differ in size. 
We attribute this to the distortion of the correlation function 
caused by the Coulomb interaction between each of the pions 
and the rest of the system \cite{bay96}. 

The parameter $R_{\rm side}$ is most directly related 
to the transverse size of the pion source. 
In order to compare it to the projectile size we first translated the 
r.m.s. radius of a gold nucleus $R_{\rm rms}$=5.3$\pm$0.1~fm \cite{preston}  
to an equivalent Gaussian radius $R_G=1/\sqrt{3} R_{\rm rms}=3.08\pm0.06$~fm. 
The ratio $R_{\rm side}/R_G$
is 1.3$\pm$0.3 for $\pi^+$ and 1.8$\pm$0.2 for $\pi^-$. 
This increase of the transverse size, 
combined with the recent finding of transverse flow velocities from analysis 
of particle spectra \cite{sch93,bra95}, 
may be used to restrict the expansion dynamics. 

It has been shown recently
that the two-meson correlation function is related to 
the meson phase space density
(defined as the number of particles per unit volume $h^3$ in 
six-dimensional phase space)
at freeze-out \cite{ber94}. 
Using the results from \cite{ber94} we obtained the following formula:
\begin{equation}
\label{phase-space-density}
\langle f\rangle _{\bf p} \ = \ 
\frac{d^3n}{d^3p} \ \ \frac{\sqrt{\pi}^3 \lambda}
{R_{\rm side}\sqrt{R_{\rm out}{}^2 R_{\rm long}{}^2-R_{\rm ol}{}^4}} \ ,
\end{equation}
where $\langle f\rangle _{\bf p}$ is the phase space density of pions with momentum 
$\bf p$, averaged over the spatial coordinates, 
$d^3n/d^3p$ is the differential pion multiplicity, 
and 
$\lambda, R_{\rm out}, R_{\rm side}$, $R_{\rm long}$, and $R_{\rm ol}$ 
are the numbers obtained from 
the `out-side-long' fit to the correlation function. 
This phase space density can be compared to a theoretical prediction 
assuming local equilibrium, i.e. to the Bose-Einstein distribution 
function. 
However, pions from long-lived resonances, 
which result in $\lambda<1$ \cite{pei92,cso96}, 
decrease the mean phase space density.
This component of the source function is not thermalized 
and should be left out from the comparison. 
Since for a case without long-lived resonances 
$\lambda=1$, 
and since the actual fraction of pions coming from the core is 
$\sqrt{\lambda}$, 
the following formula describes the average phase space density 
in the core:
\begin{equation}
\label{phase-space-density-nores}
\langle f\rangle _{\bf p}^{\rm core} \ = \  
\frac{d^3n}{d^3p} \ \ \frac{\sqrt{\pi}^3 \sqrt{\lambda}}
{R_{\rm side}\sqrt{R_{\rm out}{}^2 R_{\rm long}{}^2-R_{\rm ol}{}^4}} \ .
\end{equation}
Taking the correlation parameters from Table~II 
(with the error matrix properly taken into account) 
and using experimental $d^3n/d^3p$ from Ref. \cite{lac96}, 
we calculated $\langle f\rangle _{\bf p}^{\rm core}$ for several points 
in the $(p_t,y)$ plane (Table~III). 
We assume that the correlation parameters do not change significantly 
between the points and thus the entire momentum dependence comes from 
$d^3n/d^3p$; 
a more correct approach would require a separate correlation 
analysis for every $(p_t,y)$ bin. 

In the absence of tranverse flow and neglecting spectral changes due to 
resonance decay, 
the phase space density of pions in thermal equilibrium would follow the 
Bose-Einstein distribution (for each pion species):
\begin{equation}
f^{\rm BE}({\bf p})=\frac{1}{\exp(m_t \cosh(y-y_S)/T)-1} \ 
\label{Bose-Einstein-rapdep}
\end{equation}
with the pion transverse mass $m_t$
and rapidity $y$, and the source rapidity $y_S$. 
Since transverse momentum spectra of pions and protons at these rapidities 
have similar slopes \cite{lac96,sta96}, 
the neglect of transverse flow seems to be appropriate.
Nevertheless, 
to estimate an upper limit for the influence of flow effects 
we found by Monte Carlo simulation that transverse flow with $\beta$=0.3 
results in a phase space density lower by 20\% than the one given by 
Eq.(\ref{Bose-Einstein-rapdep}). 
Deviations of the spectral shape from thermal distribution due to $\Delta$ 
decays could be handled best if the analysis was done in bins of 
$p_t$ and $y$. 
This is planned for the high-statistics data sets taken in the '94 and '95 
runs of E877. 

Under these assumptions, we can test the thermalization of pions by 
comparing the experimental phase space density to that given by Eq. 
(\ref{Bose-Einstein-rapdep}). 
The comparison is simplified by the fact that 
the experimental differential multiplicity 
can also be parametrized by the Bose-Einstein function:
\begin{equation}
\frac{d^3n}{d^3p}=
\frac{A}{\exp(m_t/T_{\rm eff})-1} 
\end{equation}
with two fit parameters: 
effective transverse temperature $T_{\rm eff}$ and normalization $A$.
We assume that $T_{\rm eff}=T/\cosh(y-y_S)$. 
Consequently, 
the comparison between 
$\langle \!f\!\rangle _{\bf p}^{\rm core}$ and $f^{\rm BE}({\bf p})$ 
is reduced to the comparison between the parameters of the two-pion 
correlation function and the normalization of the pion spectra $A$: 
\begin{equation}
\frac{\langle f\rangle _{\bf p}^{\rm core}}{f^{\rm BE}({\bf p})} = 
A \frac{\sqrt{\pi}^3 \sqrt{\lambda}}
{R_{\rm side}\sqrt{R_{\rm out}^2 R_{\rm long}^2-R_{\rm ol}^4}} \ .
\end{equation}
Since the $p_t$-spectra have a somewhat concave shape, 
$A$ depends on the range of the fit: low (high) $p_t$'s yield high (low) $A$. 
Using $A$ obtained from a fit to the entire measured transverse momentum 
spectra $0<p_t<0.6$~GeV/c in the rapidity range 3.0--3.3 and 
the correlation parameters from Table~II, 
we evaluated the experimental-to-thermal density ratio 
of 0.97$\pm$0.23 and 1.02$\pm$0.18 for $\pi^+$ and $\pi^-$, respectively. 
Thus the experimental pion phase space density is consistent 
with the presence of local equilibrium. 

Summarizing, the multi-dimensional pion source parameters have been 
determined in the Au+Au system at the AGS. 
The observed source is significantly larger than a gold nucleus. 
The extracted pion phase space density at freeze-out 
indicates local equilibrium. 

We thank G.~Bertsch for valuable discussions. 
Financial support from the USDoE, the NSF, the Canadian NSERC, 
and CNPq Brazil is gratefully acknowledged.

\begin{table}[!b]
\caption{Parameters of the fit to one-dimensional correlation functions 
$C(Q_{\rm inv})=1+\lambda \exp {}(-R\ ^2 Q_{\rm inv}{}^2)$. 
For the purpose of comparison we also quote the results obtained with 
the Gamow correction (denoted as $G$).}
\begin{tabular}{lcccc}
& \multicolumn{2}{c}{with $H_{\Delta p}$ correction}
& \multicolumn{2}{c}{with $G$ correction}\\
& $\lambda$ & $R$ (fm) & $\lambda$ & $R$ (fm) \\
\hline
$\pi^+\pi^+$ & 0.49$\pm$0.04 & 5.5$\pm$0.3 & 0.56$\pm$0.04 & 5.1$\pm$0.2\\
$\pi^-\pi^-$ & 0.51$\pm$0.03 & 6.2$\pm$0.2 & 0.62$\pm$0.03 & 5.9$\pm$0.2\\
\end{tabular}
\end{table}

\begin{table}[hbt]
\caption{Parameters of the fit to three-dimensional correlation functions 
$C(q_{\rm out},q_{\rm side},q_{\rm long})=
$
$
1+\lambda \exp(-R_{\rm out}{}^2q_{\rm out}{}^2-
R_{\rm side}{}^2 q_{\rm side}{}^2 +
-R_{\rm long}{}^2 q_{\rm long}{}^2 
$
$
-2|R_{\rm ol}| R_{\rm ol} q_{\rm out}q_{\rm long})$.
The fit results are corrected for momentum resolution. 
That is why $\lambda$'s differ from the values in Table~I.}
\begin{tabular}{lccccc}
& $\lambda$ & $R_{\rm out}$(fm) & $R_{\rm side}$(fm) & 
   $R_{\rm long}$(fm) & $R_{\rm ol}$(fm)  \\
\hline
$\pi^+\pi^+$ & 
0.62$\pm$0.06 & 5.8$\pm$0.5 & 3.9$\pm$0.8 & 5.5$\pm$0.4 & 2.4$\pm$0.7\\ 
$\pi^-\pi^-$ & 
0.62$\pm$0.05 & 6.5$\pm$0.5 & 5.6$\pm$0.7 & 5.8$\pm$0.4 & 3.7$\pm$0.8\\ 
\end{tabular}
\end{table}

\begin{table}[hbt]
\caption{Pion phase space density at freeze-out as a function of 
$p_t$ and $y$. 
The statistical errors from $d^3n/d^3p$ are about 2\%. 
The systematic error from  $d^3n/d^3p$, 
combined with the error from the correlation analysis, 
is 24\% for $\pi^+$ and 18\% for $\pi^-$. 
(This error acts on all numbers in the Table collectively.) }
\begin{tabular}{l|ccc|ccc}
& \multicolumn{3}{c|}{$\pi^+$} & \multicolumn{3}{c}{$\pi^-$}\\
$p_t$ (GeV/c) & y=3.05 & y=3.15 & y=3.25 & y=3.05 & y=3.15 & y=3.25 \\
\hline
0.05 & 0.190 & 0.167 & 0.146 & 0.158 & 0.134 & 0.115 \\
0.10 & 0.124 & 0.110 & 0.094 & 0.100 & 0.083 & 0.071 \\
0.20 & 0.039 & 0.033 & 0.027 & 0.028 & 0.022 & 0.018 \\
\end{tabular}
\end{table}

\end{document}